# On Capacity of Wireless Ad Hoc Networks with MIMO MMSE Receivers

Jing Ma, Member, IEEE and Ying Jun (Angela) Zhang, Member, IEEE

Dept. of Information Engineering, The Chinese University of Hong Kong, Hong Kong
Email: mjingq@gmail.com, yjzhang@ie.cuhk.edu.hk

*Abstract -* Widely adopted at home, business places, and hot spots, wireless ad-hoc networks are expected to provide broadband services parallel to their wired counterparts in near future. To address this need, MIMO (multiple-input-multiple-output) techniques, which are capable of offering several-fold increase in capacity, hold significant promise. Most previous work on capacity analysis of ad-hoc networks is based on an implicit assumption that each node has only one antenna. Core to the analysis therein is the characterization of a geometric area, referred to as the exclusion region, which quantizes the amount of spatial resource occupied by a link. When multiple antennas are deployed at each node, however, multiple links can transmit in the vicinity of each other simultaneously, as interference can now be suppressed by spatial signal processing. As such, a link no longer exclusively occupies a geometric area, making the concept of "exclusion region" not applicable any more. This necessitates a revisit of the fundamental understanding of capacity of MIMO ad-hoc networks.

In this paper, we investigate link-layer throughput capacity of MIMO ad-hoc networks. In contrast to previous work, the amount of spatial resource occupied by each link is characterized by the actual interference it imposes on other links, which is a function of the correlation between the spatial channels, the distance between links, as well as the detection scheme at the receivers. To calculate the link-layer capacity, we first derive the probability distribution of post-detection SINR (signal to interference and noise ratio) at a receiver. The result is then used to calculate the number of active links and the corresponding data rates that can be sustained within an area. Our analysis shows that there exists an optimal active-link density that maximizes the link-layer throughput capacity. This will serve as a guideline for the design of medium access protocols for MIMO ad-hoc networks. To the best of knowledge, this paper is the first attempt to characterize the capacity of MIMO ad-hoc networks by considering the actual PHY-layer signal and interference model. The results in this paper pave the way for further study on network-layer transport capacity of ad-hoc networks with MIMO.

# I. INTRODUCTION

MIMO (Multi-Input Multi-Output) systems where multiple antennas are deployed at both transmitter and receiver, open up a new dimension, i.e., space, to significantly improve the spectral efficiency of wireless communication systems. Foschini and Telatar [1]-[2] show that MIMO provides a linear growth of capacity with the number of antennas. Moreover, the extra degree of freedom, i.e., space, offered by multiple antennas enables interference cancelation at receiving stations, which allows spectrum to be reused more aggressively [3]-[7]. On the other hand, MANET (mobile ad hoc network) is likely to play a major role in next-generation home networks and hot spots, thanks to its simplicity, cost effectiveness, and simple reconfiguration [8]. One of the major challenges faced by MANET is the increasing demand for data-rate-intensive applications similar to those in its wired counterpart. Deploying multiple antennas at each node is a promising solution to improve network capacity and link reliability required by these applications. To fully exploit the benefits of MIMO in ad hoc networks, it is essential to have a thorough understanding of the fundamental impact of the use of MIMO on overall network performance.

Most previous work on capacity analysis of ad hoc networks is based on an implicit assumption that only one antenna is used at each node [9]. Under this assumption, an active link exclusively occupies a geometric area, referred to as the exclusion region, to avoid collisions with other links. The larger the exclusion region, the more spatial resource a link occupies, and the capacity of a network is essentially determined by the amount of spatial resource occupied by underlying links. When multiple antennas are deployed at each node, however, an active link no longer exclusively occupies a spatial area, making the original definition of "exclusion region" not applicable any more. In fact, multiple adjacent links can transmit at the same time, as long as the mutual interference can be suppressed by spatial signal processing. As such, previous work on capacity analysis is not directly applicable to ad hoc networks with MIMO links.

The impact of multiple antennas on network capacity was previously studied under different contexts. In [23]-[24], Zhang and Liew derived the upper and lower bounds on network capacity when directional antennas with realizable generic patterns are used. It is well known that directional antennas do not work well in indoor and urban environments where a large number of local scatterers cause severe multipath effects and large angular spread. Unfortunately, most application scenarios of ad hoc networks perceive rich-scattering channels. As a result, the previous analysis based on directional antennas does not apply to general MIMO ad hoc networks. In [15], Chen et al studied the ergodic single-hop capacity of ad-hoc networks when single-user detection is employed at each receiver. Though simple, single-user detectors do not exploit the interference cancelation capability of MIMO. As a result, the capacity calculated in [15] is far below the actual achievable capacity of MIMO ad hoc networks.

In this paper, we investigate link-layer throughput capacity, defined as the total data rate that can be successfully delivered through all single-hop links within a unit area, of MIMO ad-hoc networks. In particular, MMSE (minimum mean square error), the optimal linear multiuser detection, is assumed to be deployed at receiving nodes, as it has the highest interference suppression capability among all linear detection schemes including ZF (zero forcing) and single user detection [25]. In contrast to previous work, we characterize the amount of spatial resource occupied by each link by the actual interference active links impose on each other, taking into consideration the actual behavior of multipath fading and MIMO systems.

One of the key challenges in this work is the characterization of the distribution of post-detection SINR (signal to interference and noise ratio) when transmitters are randomly located. SINR distribution of MMSE detectors was previously studied in [26]-[28], [19]. In [27] and [28], the authors proved asymptotic Normality of post-detection SINR for equal and non-equal interference power, respectively. In [19], Li et al improved the accuracy by modeling SINR using a Gamma or a generalized Gamma distribution. All these papers assumed that the signal strength of interfering data streams as detected at the receiver is deterministic and known. In ad-hoc networks, however, active nodes are randomly located, and hence the received interference power is also random. Moreover, previous work often assumed that the number of interferers is smaller than or comparable to the number of receive antennas. While the assumption is reasonable for traditional cellular networks, it is not true in ad hoc networks where the number of simultaneously transmitting stations could be much larger than the number of antennas at a receiving node. The main contributions of this paper can be summarized as follows.

- We derive a closed-form expression for the distribution of received SINR of an active link with maximum ratio transmission and a multiuser MMSE receiver. In contrast to [26]-[28], [19], signal power, interference power, and the number of interferers are all random variables due to the randomness in the location and activity of transmitting nodes. The analytical results are validated by numerical simulations.

- Based on the SINR distribution, we analyze link-layer throughput capacity, calculated as the sum data rate that can be delivered by all links within a unit area. In contrast to [15] where the data rate of each link is represented by ergodic capacity, per-link data rate is defined as $Th = (1-P_{out})q$ in this paper, where $q$ is the transmission rate of each link and $P_{out}$ is the probability of transmission failure given $q$. This definition more accurately reflects the characteristics of today's ad-hoc networks, where multipath fading varies slowly compared with the transmission time of a packet. Unlike the mean value analysis in [19], the PDF (probability density function) of SINR is needed in this paper to calculate $P_{out}$, which makes the job here much more challenging. Our work on link-layer capacity paves the way for further study on network-layer transport capacity of ad-hoc networks with MIMO links.

- The analysis suggests that there exists an optimal density of simultaneously transmitting links that maximizes the link-layer capacity. Through numerical study, we calculate the optimal density under various scenarios. In real implementation, the optimal active-link density can be mapped to an optimal transmission probability in the MAC (medium access control) layer. This result serves as a guideline for the design of MAC protocols of next-generation ad-hoc networks with MIMO links.

- Our analysis is based on the assumption that CSI (channel state information) from all transmitting nodes to the receiver of a tagged link is available at the receiver. In practice, it is likely that a receiver only knows CSI from a few neighboring links. In this paper, we also study the link-layer capacity and the corresponding optimal link density when only local CSI is available.

The remainder of the paper is organized as follows. The system and signal models are presented in Section II. In Section III, we derive the distribution of post-detection SINR in ad hoc networks when MMSE receivers are deployed. The link-layer throughput capacity is then given in Section IV. Numerical examples and discussions are presented in Section V, where we also study the impact of partial CSI on the capacity and optimal link density. Finally, the paper is concluded in Section VI.

## II. SYSTEM MODEL

We first describe the notation used in this paper for readers' convenience. Throughout the paper, scalars are given by normal letters, vectors by boldface lower case letters, and matrices by boldface upper case letters. Besides, the following notations are used.

$\mathbf{X}^T$ : Transpose operation
$\mathbf{X}^*$ : Hermitian transpose
$[\mathbf{X}]_{ij}$ : $(i, j)$ th element of $\mathbf{X}$
$\mathrm{Tr}(\mathbf{X})$ : trace of matrix $\mathbf{X}$

$E(\cdot)$: Expectation
$Var(\cdot)$: Variance

*A. System Description*

Consider an ad hoc network as demonstrated in Fig. 1, where mobile nodes are uniformly distributed within an area. Each node is equipped with $m$ antennas. For simplicity, we ignore the edge effects and assume that each link has the same statistical characteristics. Without loss of generality, let Link 0 be the tagged link. At a given time, there are $K$ other links sending data at the same time as Link 0, resulting in co-channel interference. As a result, the data received by the tagged link, given as follows, is a superposition of desired signal, interference, and noise.

$$\mathbf{y}_0 = \sqrt{\alpha_0 p_0}\mathbf{H}_0\mathbf{x}_0 + \sum_{k=1}^{K}\sqrt{\alpha_k p_k}\mathbf{H}_k\mathbf{x}_k + \mathbf{n}_0 \qquad (1)$$

In the above, $\mathbf{H}_k$ is an $m \times m$ channel matrix, representing the channel fading from the transmitter of the $k$ th link to the receiver of the *tagged* link. Assuming a rich scattering environment and quasi-static Rayleigh flat fading channels, we can model the elements of $\mathbf{H}_k$ as i.i.d. complex Gaussian random variables. Likewise, $\mathbf{x}_0$ denotes the transmit signal vector of Link $k$; $p_k$ the transmit power of Link $k$; $\alpha_k$ the path loss from the transmitter of Link $k$ to the receiver of the tagged link; and $\mathbf{n}_0$ the AWGN (additive white Gaussian noise) with zero mean and unit variance.

Note that the number of interferers, $K$, is a random variable depending on the transmission probability of links. Since the neighborhood observed by each link is statistically identical, we assume that the $K$ interferers are randomly located within a disc of radius $R$ centered at the tagged receiver, where $R$ is the largest distance at which an interferer can cause non-negligible interference to the receiver. Furthermore, let $\varepsilon$ denote the minimum separation between interferers and the tagged receiver. Assume that $\varepsilon$ is small enough so that it does not affect the uniform distribution of nodes. Thus, the probability density function of the distance between a node and the tagged receiver is given by

$$f_d(x) = \frac{2x}{R^2 - \varepsilon^2} \qquad (2)$$

Let $c_k$ be the distance between the $k$ th transmitter to the tagged receiver and $\theta$ be the path loss exponent. In particular, $c_0$ is the length of the tagged link. We can calculate the received power from the $k$ th interferer as

$$\alpha_k p_k = \left(\frac{c_0}{c_k}\right)^\theta \alpha_0 p_0 \tag{3}$$

whose PDF is

$$f_{\alpha_k p_k}(x) = \frac{2(\alpha_0 p_0)^{2/\theta} c_0^2}{\theta(R^2 - \varepsilon^2) x^{(\theta+2)/\theta}} \quad \forall \left(\frac{c_0}{R}\right)^\theta \alpha_0 p_0 \leq x \leq \left(\frac{c_0}{\varepsilon}\right)^\theta \alpha_0 p_0. \tag{4}$$

When $\theta = 4$,

$$f_{\alpha_k p_k}(x) = \frac{\sqrt{\alpha_0 p_0} c_0^2}{2(R^2 - \varepsilon^2) x^{3/2}} \quad \forall \left(\frac{c_0}{R}\right)^4 \alpha_0 p_0 \leq x \leq \left(\frac{c_0}{\varepsilon}\right)^4 \alpha_0 p_0 \tag{5}$$

*B. Maximum Ratio Transmission and MMSE Reception*

It was proved in [29] that SVD (singular value decomposition) based space-time vector coding allows the collection of signal power in space and it is a theoretical means to achieve high capacity for MIMO systems. By SVD, $\mathbf{H}_k$ can be decomposed into

$$\mathbf{H}_k = \sum_{j=1}^{r_k} \lambda_{k,j} \mathbf{u}_{k,j} \mathbf{v}_{k,j}^*, \tag{6}$$

where $\lambda_{k,1} \geq \lambda_{k,2} \geq \cdots \geq \lambda_{k,r_k}$ are the eigenvalues, $\mathbf{u}_{k,j}$ and $\mathbf{v}_{k,j}$ are the left singular vector and right singular vector, respectively, and $r_k$ is the rank of $\mathbf{H}_k$. Note that the left and right singular vectors have the same distribution as normalized complex Gaussian random vectors [20]-[21]. Likewise, the distribution of the square of the largest singular value $\lambda_{k,1}^2$ is given by [17] as a finite linear combination of elementary Gamma densities:

$$f_{\lambda_{k,1}^2}(x) = \sum_{n=1}^{m} \sum_{l=0}^{2mn-2n^2} g_{n,l} \frac{n^{l+1} x^l e^{-nx}}{l!} \quad \forall x > 0, \tag{7}$$

where $g_{n,l}$ are computed and listed in [17] for most antenna configurations of interest. The $\tau$ th moment of $\lambda_{k,1}^2$ is

$$\mathrm{E}\left([\lambda_{k,1}^2]^\tau\right) = \sum_{n=1}^{m} \sum_{l=0}^{2mn-2n^2} \frac{g_{n,l}(l+\tau)!}{n^\tau l!}. \tag{8}$$

In an interference-limited environment such as ad hoc networks, an active link should transmit only one data stream at a time to optimize the system performance [11]-[13]. In this case, the single data stream should be transmitted on the largest singular mode of the channel for SNR (signal to noise ratio) maximization. Such scheme, known as MRT (maximum ratio transmission), configures the transmit antenna weight using the right singular vector corresponding to the dominant singular value.[1] For example, the transmit antenna weight of Link 0 is $\mathbf{v}_{0,1}$. Similarly, the transmit beamforming vector of Link $k$, denoted by $\mathbf{t}_k$, is the dominant singular vector of the channel matrix between its own transmitter-receiver pair. Therefore, the received signal in (1) becomes

$$\mathbf{y}_0 = \sqrt{\alpha_0 p_0} \mathbf{H}_0 \mathbf{v}_{0,1} b_0 + \sum_{k=1}^{K} \sqrt{\alpha_k p_k} \mathbf{H}_k \mathbf{t}_k b_k + \mathbf{n}_0$$

$$= \sqrt{\alpha_0 p_0} \lambda_{0,1} \mathbf{u}_{0,1} b_0 + \sum_{k=1}^{K} \sqrt{\alpha_k p_k} \hat{\mathbf{h}}_k b_k + \mathbf{n}_0 \qquad (9)$$

where $\mathbf{y}_0$ is a $m \times 1$ vector with the $i$th element being the received signal on the $i$th receive antenna. Denote $\mathbf{H}_k \mathbf{t}_k$ by $\hat{\mathbf{h}}_k$, whose elements are still i.i.d. complex Gaussian random variables with zero mean and unit variance [2], since $\mathbf{t}_k$ has unit norm and is independent of $\mathbf{H}_k$.

Define equivalent channel matrix $\mathbf{G}$ as

$$\mathbf{G} = [\lambda_{0,1} \mathbf{u}_{0,1}, \hat{\mathbf{h}}_1, \cdots, \hat{\mathbf{h}}_K], \qquad (10)$$

and the transmit power matrix as

$$\mathbf{P} = \begin{bmatrix} \alpha_0 p_0 & & & \\ & \alpha_1 p_1 & & \\ & & \ddots & \\ & & & \alpha_K p_K \end{bmatrix}. \qquad (11)$$

We can then rewrite (9) into a matrix form as

$$\mathbf{y}_0 = \mathbf{G}\mathbf{P}^{1/2}\mathbf{b} + \mathbf{n}_0, \qquad (12)$$

where $\mathbf{b}$ is $[b_0, b_1, \cdots, b_K]^T$.

Upon receiving the signal, the tagged receiver attempts to obtain an estimate of $\mathbf{b}$ from the received signal $\mathbf{y}_0$. Being the optimal linear detector, MMSE detector minimizes the mean square error between $\mathbf{b}$ and its estimate. Specifically, the decision statistics $\tilde{\mathbf{b}}$ is obtained by linearly combining the received signal vector as follows:

---

[1] To implement MRT, transmitter-side CSI is needed. Transmitter-side CSI is easily achievable in wireless networks with two-way communications. In case it is not available, random antenna selection or space time coding can be deployed instead of MRT. Our analysis can be easily extended to these cases with slight modification. Specifically, it is the distribution of $\lambda_{0,1}$ that needs to be modified in the analysis.

$$\tilde{\mathbf{b}} = \mathbf{V}^*\mathbf{y}_0 \qquad (13)$$

where $\mathbf{V} = \left(\mathbf{I}_{K+1} + \mathbf{G}^*\mathbf{P}\mathbf{G}\right)^{-1}\mathbf{G}\mathbf{P}$ and $\mathbf{I}_{K+1}$ is a $(K+1)\times(K+1)$ Identity matrix. With MMSE, the post-detection SINR of the tagged link can be calculated as [25]

$$\text{SINR}_0 = \frac{1}{\left[\left(\mathbf{I}_{K+1} + \mathbf{G}^*\mathbf{P}\mathbf{G}\right)^{-1}\right]_{1,1}} - 1 \qquad (14)$$

where $[\cdot]_{1,1}$ denotes the $(1,1)^{th}$ element of a matrix. The distribution of SINR was previously studied in [26]-[28], [19]. However, their work assumes that interference power (i.e., $\alpha_k p_k$ for $k > 0$) is deterministic and known. This assumption, however, is not applicable to ad hoc networks where interfering links are randomly located. In this paper, we focus on MMSE receivers, for it achieves the optimal performance in terms of BER (bit error rate) or SINR among all linear detectors. Our conclusions, however, can easily be extended to other suboptimal detectors such as ZF (zero forcing) and single-user detection.

Note that eqns. (13) and (14) have assumed that the tagged receiver has the knowledge of $\hat{\mathbf{h}}_k$ for all $k$. Although not realistic, this assumption allows us to investigate the fundamental limit of wireless MIMO network capacity without taking into account implementation details. This assumption will later be removed in Section V in Fig. 8 and Fig. 9, where the receiver only knows the CSI from its neighboring interfering nodes.

## III. SINR DISTRIBUTION OF MMSE

In this section, we derive the distribution of SINR in ad hoc networks when MMSE detection is deployed. To this end, we first compute the mean and variance of SINR in subsections III.A to III.E. The PDF of SINR is then presented in III.F.

*A. Simplified Form of SINR*

We define

$$\begin{aligned}\tilde{\mathbf{G}} &= \mathbf{P}^{1/2}\mathbf{G} \\ &= \left[\sqrt{\alpha_0 p_0}\lambda_{0,1}\mathbf{u}_{0,1}, \sqrt{\alpha_1 p_1}\hat{\mathbf{h}}_1, \ldots, \sqrt{\alpha_K p_K}\hat{\mathbf{h}}_K\right].\end{aligned} \qquad (15)$$

and then have

$$\tilde{\mathbf{G}}^*\tilde{\mathbf{G}} = \begin{bmatrix} \alpha_0 p_0 \lambda_{0,1}^2 & \sqrt{\alpha_0 p_0}\lambda_{0,1}\mathbf{u}_{0,1}^*\tilde{\mathbf{G}}_{-1} \\ \tilde{\mathbf{G}}_{-1}^*\lambda_{0,1}\mathbf{u}_{0,1} & \tilde{\mathbf{G}}_{-1}^*\tilde{\mathbf{G}}_{-1} \end{bmatrix}, \tag{16}$$

where $\tilde{\mathbf{G}}_{-1}$ is $\tilde{\mathbf{G}}$ with first column removed.

Before going further, we first describe the following lemma.

***Lemma 1***: *Write a matrix* $\mathbf{A}$ *into*

$$\mathbf{A} = \begin{bmatrix} a_{1,1} & a_{1,-1}^* \\ a_{1,-1} & \mathbf{A}_{-1,-1} \end{bmatrix}, \tag{17}$$

where $a_{1,1}$ is the $(1,1)^{th}$ element of $\mathbf{A}$, $a_{1,-1}$ is the first column of $\mathbf{A}$ with the first element removed and $\mathbf{A}_{-1,-1}$ is $\mathbf{A}$ with the first column and row removed. Then,

$$[\mathbf{A}^{-1}]_{1,1} = \left(a_{1,1} - a_{1,-1}^*(\mathbf{A}_{-1,-1})^{-1}a_{1,-1}\right)^{-1}. \tag{18}$$

By using Lemma 1, (14) can be simplified as

$$\text{SINR}_0 = \frac{1}{\left[\left(\mathbf{I}_{K+1} + \tilde{\mathbf{G}}^*\tilde{\mathbf{G}}\right)^{-1}\right]_{1,1}} - 1$$

$$= \alpha_0 p_0 \lambda_{0,1}^2 - \alpha_0 p_0 \lambda_{0,1}^2 \mathbf{u}_{0,1}^*\tilde{\mathbf{G}}_{-1}\left(\mathbf{I}_K + \tilde{\mathbf{G}}_{-1}^*\tilde{\mathbf{G}}_{-1}\right)^{-1}\tilde{\mathbf{G}}_{-1}^*\mathbf{u}_{0,1}. \tag{19}$$

Denoting the SVD of $\tilde{\mathbf{G}}_{-1}$ as

$$\tilde{\mathbf{G}}_{-1} = \mathbf{WDZ}, \tag{20}$$

where the $i$ th diagonal element of $\mathbf{D}$ is $d_i$, we then can derive the SINR as

$$\text{SINR}_0 = \alpha_0 p_0 \lambda_{0,1}^2 - \alpha_0 p_0 \lambda_{0,1}^2 \mathbf{u}_{0,1}^*\mathbf{WDZ}\left(\mathbf{I}_K + \mathbf{Z}^*\mathbf{D}^*\mathbf{DZ}\right)^{-1}\mathbf{Z}^*\mathbf{D}^*\mathbf{W}^*\mathbf{u}_{0,1}$$

$$= \alpha_0 p_0 \lambda_{0,1}^2 \mathbf{u}_{0,1}^*\mathbf{W}\left(\mathbf{I}_m - \left(\mathbf{I}_m + \mathbf{D}^{*-1}\mathbf{D}^{-1}\right)^{-1}\right)\mathbf{W}^*\mathbf{u}_{0,1}$$

$$= \alpha_0 p_0 \lambda_{0,1}^2 \mathbf{u}_{0,1}^*\mathbf{W}(\mathbf{I}_m + \mathbf{DD}^*)^{-1}\mathbf{W}^*\mathbf{u}_{0,1}$$

$$= \alpha_0 p_0 \lambda_{0,1}^2 \mathbf{u}_{0,1}^*\mathbf{B}\mathbf{u}_{0,1}, \tag{21}$$

where $\mathbf{B}$ is defined as $\mathbf{W}(\mathbf{I}_m + \mathbf{DD}^*)^{-1}\mathbf{W}^*$.

*B. Conditional Mean of SINR*

It is easy to see that $\mathbf{B}$ is deterministic function of $\mathbf{G}$ and $\mathbf{P}$. Given a channel realization, the conditional mean of $\text{SINR}_0$ given $\mathbf{B}$ is

$$E(\text{SINR}_0 \mid \mathbf{B}) = \alpha_0 p_0 E(\lambda_{0,1}^2) E(\mathbf{u}_{0,1}^* \mathbf{B} \mathbf{u}_{0,1}). \tag{22}$$

Likewise,

$$E(\mathbf{u}_{0,1}^* \mathbf{B} \mathbf{u}_{0,1}) = E\left(\sum_{i=1}^m u_{0,1}^{*(i)} u_{0,1}^{(i)} B_{ii}\right) + E\left(\sum_{i \neq j} u_{0,1}^{*(i)} u_{0,1}^{(j)} B_{ij}\right)$$

$$= \sum_{i=1}^m E\left(|\mathbf{u}_{0,1}^{(i)}|^2\right) B_{ii} + \sum_{i \neq j} E\left(u_{0,1}^{*(i)} u_{0,1}^{(j)}\right) B_{ij}$$

$$= \frac{1}{m} \text{Tr}(\mathbf{B}) + \sum_{i \neq j} E\left(u_{0,1}^{*(i)} u_{0,1}^{(j)}\right) B_{ij} \tag{23}$$

where $B_{ij}$ is the $(i,j)$ th element of $\mathbf{B}$, and $u_{0,1}^{(i)}$ is the $i$ th element of $\mathbf{u}_{0,1}$. Since $\mathbf{u}_{0,1}$ is a normalized complex Gaussian random vector as mentioned in Section II, it is not difficult to prove that

$$E\left(u_{0,1}^{*(i)} u_{0,1}^{(j)}\right) = 0 \; \forall i \neq j. \tag{24}$$

Hence, we have the conditional expectation of $\text{SINR}_0$ given $\mathbf{B}$ as shown in (25).

$$E(\text{SINR}_0 \mid \mathbf{B}) = \frac{1}{m} \alpha_0 p_0 E(\lambda_{0,1}^2) \text{Tr}\left(\mathbf{W}\left(\mathbf{I}_m + \mathbf{D}\mathbf{D}^*\right)^{-1} \mathbf{W}^*\right)$$

$$= \frac{1}{m} \alpha_0 p_0 E(\lambda_{0,1}^2) \text{Tr}\left(\left(\mathbf{I}_m + \mathbf{D}\mathbf{D}^*\right)^{-1}\right)$$

$$= \alpha_0 p_0 E(\lambda_{0,1}^2) \left(\frac{1}{m} \sum_{i=1}^m \frac{1}{1+d_i^2}\right) \tag{25}$$

## C. Conditional Second Moment of SINR

We now derive the conditional second moment of SINR given $\mathbf{B}$.

Denoting $E(|u_{0,1}^{(i)}|^4) \, (1 \leq i \leq m)$ as $a_1$, and $E(|u_{0,1}^{(i)}|^2 |u_{0,1}^{(j)}|^2) \, (i \neq j)$ as $a_2$, respectively, we first derive (26) from (23).

$$E\left((\mathbf{u}_{0,1}^* \mathbf{B} \mathbf{u}_{0,1})^2\right) = \sum_{i=1}^m E(|u_{0,1}^{(i)}|^4) B_{ii}^* B_{ii} + \sum_{i \neq j} E(|u_{0,1}^{(i)}|^2 |u_{0,1}^{(j)}|^2) B_{ij}^* B_{ij} + \sum_{i \neq j} E(|u_{0,1}^{(i)}|^2 |u_{0,1}^{(j)}|^2) B_{ii}^* B_{jj}$$

$$+ \sum_{i_1 \neq j_1, i_2 \neq j_2, i_1 \neq i_2, j_1 \neq j_2} E(u_{0,1}^{(i_1)} u_{0,1}^{(j_2)} u_{0,1}^{*(j_1)} u_{0,1}^{*(i_2)}) B_{i_1 j_1}^* B_{i_2 j_2}^*$$

$$= \sum_{i=1}^m E(|u_{0,1}^{(i)}|^4) B_{ii}^* B_{ii} + \sum_{i \neq j} E(|u_{0,1}^{(i)}|^2 |u_{0,1}^{(j)}|^2) B_{ij}^* B_{ij} + \sum_{i \neq j} E(|u_{0,1}^{(i)}|^2 |u_{0,1}^{(j)}|^2) B_{ii}^* B_{jj}$$

$$= a_1 \sum_{i=1}^m B_{ii}^* B_{ii} + a_2 \text{Tr}(\mathbf{B}^* \mathbf{B}) + a_2 \text{Tr}^*(\mathbf{B}) \text{Tr}(\mathbf{B}) - 2 a_2 \sum_{i=1}^m B_{ii}^* B_{ii}$$

$$= a_2\left(\text{Tr}(\mathbf{B}^*\mathbf{B}) + \text{Tr}^*(\mathbf{B})\text{Tr}(\mathbf{B})\right) + (a_1 - 2a_2)\sum_{i=1}^{m} B_{ii}^* B_{jj} \tag{26}$$

where the second equality is due to the fact that

$$\text{E}\left(u_{0,1}^{(i_1)} u_{0,1}^{(j_2)} u_{0,1}^{*(j_1)} u_{0,1}^{*(i_2)}\right) = 0 \ \forall i_1 \neq j_1, i_2 \neq j_2, i_1 \neq i_2, j_1 \neq j_2 \tag{27}$$

Since $\mathbf{u}_{0,1}$ is a normalized complex Gaussian random vector, the PDF of $|u_{0,1}^{(i)}|^2$ is

$$f_{|u_{0,1}^{(i)}|^2}(x) = (m-1)(1-x)^{m-2}, 0 \leq x \leq 1. \tag{28}$$

Therefore, we have

$$\text{E}(|u_{0,1}^{(i)}|^2) = \frac{1}{m}, \tag{29}$$

and

$$\text{E}(|u_{0,1}^{(i)}|^4) = \frac{2}{m(m+1)}. \tag{30}$$

We can then derive the pdf of $|u_{0,1}^{(i)}|^2$ on the condition of $|u_{0,1}^{(j)}|^2 = y$ as

$$f_{|u_{0,1}^{(i)}|^2 \big| |u_{0,1}^{(j)}|^2}(x|y) = \frac{m-2}{1-y}\left(\frac{1-x-y}{1-y}\right)^{m-3} \quad 0 \leq x \leq 1-y. \tag{31}$$

Then, we have $\text{E}(|u_{0,1}^{(i)}|^2 |u_{0,1}^{(j)}|^2)$ as follows.

$$\text{E}(|u_{0,1}^{(i)}|^2 |u_{0,1}^{(j)}|^2) = \int_0^1 \int_0^{1-y} xy f_{|u_{0,1}^{(i)}|^2 \big| |u_{0,1}^{(j)}|^2}(x|y) f_{|u_{0,1}^{(j)}|^2}(y) dx dy$$

$$= \int_0^1 y(m-1)(1-y)^{m-2} \int_0^{1-y} x \frac{m-2}{1-y}\left(\frac{1-x-y}{1-y}\right)^{m-3} dx dy$$

$$= \frac{1}{m(m+1)} \tag{32}$$

From eqns. (26) and (32), we have

$$\text{E}\left((\mathbf{u}_{0,1}^*\mathbf{B}\mathbf{u}_{0,1})^2\right) = \frac{1}{m(m+1)}\left(\text{Tr}(\mathbf{B}^*\mathbf{B}) + \text{Tr}^*(\mathbf{B})\text{Tr}(\mathbf{B})\right)$$

$$= \frac{1}{m(m+1)}\left(\sum_{i=1}^{m}\left(\frac{1}{1+d_i^2}\right)^2 + \left(\sum_{i=1}^{m}\frac{1}{1+d_i^2}\right)^2\right) \tag{33}$$

We are now ready to derive the second moment of $\text{SINR}_0$ as

$$\mathrm{E}(\mathrm{SINR}_0^2 \mid \mathbf{B}) = \alpha_0^2 p_0^2 \mathrm{E}(\lambda_{0,1}^4) \mathrm{E}\left((\mathbf{u}_{0,1}^* \mathbf{B} \mathbf{u}_{0,1})^2\right)$$

$$= \alpha_0^2 p_0^2 \mathrm{E}(\lambda_{0,1}^4) \frac{1}{m(m+1)} \left( \left(\sum_{i=1}^m \frac{1}{1+d_i^2}\right)^2 + \sum_{i=1}^m \left(\frac{1}{1+d_i^2}\right)^2 \right). \quad (34)$$

*D. Mean of SINR*

In order to compute the mean of $\mathrm{SINR}_0$, we first introduce the method of asymptotic analysis of random matrix [18].

**Definition of $\eta$ transform:**

Given a nonnegative random variable $\chi$, the $\eta$ transform is defined as

$$\eta_\chi(\gamma) = \mathrm{E}\left(\frac{1}{1+\gamma\chi}\right) \quad (35)$$

***Theorem 1*** [18]*:* Let $\mathbf{H}$ be an $m \times K$ matrix whose entries are i.i.d. complex Gaussian variables with variance $\frac{1}{m}$. Let $\mathbf{T}$ be a $K \times K$ Hermitian nonnegative random matrix, independent of $\mathbf{H}$, whose empirical eigenvalue distribution converges almost surely to a nonrandom limit. The empirical eigenvalue distribution of $\mathbf{HTH}^*$ converges almost surely, as $K, m \to \infty$ with $\frac{K}{m} \to \beta$, to a distribution whose $\eta$-transform satisfies

$$\beta = \frac{1-\eta}{1-\eta_{\mathbf{T}}(\gamma\eta)}, \quad (36)$$

where for simplicity we have abbreviated $\eta_{\mathbf{HTH}^*}(\gamma) = \eta$. $\eta_{\mathbf{HTH}^*}(\cdot)$ and $\eta_{\mathbf{T}}(\cdot)$ stand for the $\eta$ transform of the eigenvalues of $\mathbf{HTH}^*$ and $\mathbf{T}$, respectively.

Theorem 1 can be applied to find the empirical eigenvalue distribution of $\tilde{\mathbf{G}}_{-1}\tilde{\mathbf{G}}_{-1}^*$, i.e., the empirical distribution of $d_i^2$. Rewrite the matrix as

$$\tilde{\mathbf{G}}_{-1}\tilde{\mathbf{G}}_{-1}^* = \frac{1}{\sqrt{m}}\mathbf{G}_{-1} m \mathbf{P}_{-1} \frac{1}{\sqrt{m}}\mathbf{G}_{-1}^*. \quad (37)$$

where $\mathbf{P}_{-1}$ is $\mathbf{P}$ with first column removed. Then, from Theorem 1, the empirical eigenvalue distribution of $\tilde{\mathbf{G}}_{-1}\tilde{\mathbf{G}}_{-1}^*$ converges almost surely to a distribution whose $\eta$ transform satisfies

$$\frac{K}{m} = \frac{1-\eta_{\tilde{\mathbf{G}}_{-1}\tilde{\mathbf{G}}_{-1}^*}(\gamma)}{1-\eta_{m\mathbf{P}_{-1}}(\gamma\eta_{\tilde{\mathbf{G}}_{-1}\tilde{\mathbf{G}}_{-1}^*}(\gamma))}, \tag{38}$$

where $\eta_{m\mathbf{P}_{-1}}(\gamma)$ is the $\eta$ transform of the eigenvalue of $m\mathbf{P}_{-1}$.

We now begin to derive $\eta_{m\mathbf{P}_{-1}}(\gamma)$. Since $\mathbf{P}_{-1}$ is a diagonal matrix, the empirical eigenvalue distribution of $\mathbf{P}_{-1}$ is the distribution of its diagonal elements of $\alpha_k p_k$. Given the distribution of $\alpha_k p_k$ in (3), we derive the $\eta$ transform of $m\mathbf{P}_{-1}$ as

$$\eta_{m\mathbf{P}_{-1}}(\gamma) = \mathrm{E}\left(\frac{1}{1+\gamma m\alpha_k p_k}\right)$$

$$= 1 - \frac{c_0^2\sqrt{m\alpha_0 p_0\gamma}}{R^2-\varepsilon^2}\left(\tan^{-1}\frac{c_0^2\sqrt{m\alpha_0 p_0\gamma}}{\varepsilon^2} - \tan^{-1}\frac{c_0^2\sqrt{m\alpha_0 p_0\gamma}}{R^2}\right) \tag{39}$$

Substituting (39) to (38), we have

$$1-\eta_{\tilde{\mathbf{G}}_{-1}\tilde{\mathbf{G}}_{-1}^*}(\gamma) = \frac{Kc_0^2\sqrt{m\alpha_0 p_0\gamma\eta_{\tilde{\mathbf{G}}_{-1}\tilde{\mathbf{G}}_{-1}^*}(\gamma)}}{m(R^2-\varepsilon^2)}(\tan^{-1}(\frac{c_0^2\sqrt{m\alpha_0 p_0\gamma\eta_{\tilde{\mathbf{G}}_{-1}\tilde{\mathbf{G}}_{-1}^*}(\gamma)}}{\varepsilon^2})$$

$$-\tan^{-1}(\frac{c_0^2\sqrt{m\alpha_0 p_0\gamma\eta_{\tilde{\mathbf{G}}_{-1}\tilde{\mathbf{G}}_{-1}^*}(\gamma)}}{R^2})). \tag{40}$$

We are now ready to derive the mean of $\mathrm{SINR}_0$ as

$$\mathrm{E}(\mathrm{SINR}_0) = \mathrm{E}_{\mathbf{B}}\left(\mathrm{E}(\mathrm{SINR}_0\mid\mathbf{B})\right)$$

$$= \alpha_0 p_0 \mathrm{E}(\lambda_{0,1}^2)\mathrm{E}\left(\frac{1}{m}\sum_{i=1}^{m}\frac{1}{1+d_i^2}\right). \tag{41}$$

Note that

$$\mathrm{E}\left(\frac{1}{m}\sum_{i=1}^{m}\frac{1}{1+d_i^2}\right) = \eta_{\tilde{\mathbf{G}}_{-1}\tilde{\mathbf{G}}_{-1}^*}(1), \tag{42}$$

which can be computed by (40) as

$$1-\eta_{\tilde{\mathbf{G}}_{-1}\tilde{\mathbf{G}}_{-1}^*}(1) = \frac{Kc_0^2\sqrt{m\alpha_0 p_0\eta_{\tilde{\mathbf{G}}_{-1}\tilde{\mathbf{G}}_{-1}^*}(1)}}{m(R^2-\varepsilon^2)}(\tan^{-1}(\frac{c_0^2\sqrt{m\alpha_0 p_0\eta_{\tilde{\mathbf{G}}_{-1}\tilde{\mathbf{G}}_{-1}^*}(1)}}{\varepsilon^2})$$

$$-\tan^{-1}(\frac{c_0^2\sqrt{m\alpha_0 p_0 \eta_{\tilde{G}_{-1}\tilde{G}_{-1}^*}(1)}}{R^2})). \tag{43}$$

At last, we have the mean of $SINR_0$ as

$$E(SINR_0) = \alpha_0 p_0 E(\lambda_{0,1}^2) E\left(\frac{1}{m}\sum_{i=1}^{m}\frac{1}{1+d_i^2}\right)$$

$$= \alpha_0 p_0 E(\lambda_{0,1}^2) \eta_{\tilde{G}_{-1}\tilde{G}_{-1}^*}(1) \tag{44}$$

*E. Variance of SINR*

We now begin to derive the various of $SINR_0$.

$$Var(SINR_0) = E_\mathbf{B}(Var(SINR_0|\mathbf{B})) + Var(E(SINR_0|\mathbf{B}))$$

$$\approx E(Var(SINR_0|\mathbf{B})), \tag{45}$$

where the approximation is due to the fact that $Var(E(SINR_0|\mathbf{B}))$ converges to zero as the rank of $\mathbf{B}$ becomes very large.

$$Var(SINR_0) \approx E_\mathbf{B}(Var(SINR_0|\mathbf{B}))$$

$$= E_\mathbf{B}\left(E(SINR_0^2|\mathbf{B}) - E^2(SINR_0|\mathbf{B})\right)$$

$$= \alpha_0^2 p_0^2 E_\mathbf{B}\left\{E(\lambda_{0,1}^4)\frac{1}{m(m+1)}\left[\sum_{i=1}^{m}\left(\frac{1}{1+d_i^2}\right)^2 + \left(\sum_{i=1}^{m}\frac{1}{1+d_i^2}\right)^2\right]\right.$$

$$\left.-E^2(\lambda_{0,1}^2)\left(\frac{1}{m}\sum_{i=1}^{m}\frac{1}{1+d_i^2}\right)^2\right\}$$

$$\stackrel{(i)}{\approx} \alpha_0^2 p_0^2 \frac{E(\lambda_{0,1}^4) - E^2(\lambda_{0,1}^2)}{m} E\left(\sum_{i=1}^{m}\left(\frac{1}{1+d_i^2}\right)^2\right)$$

$$= \alpha_0^2 p_0^2 (E(\lambda_{0,1}^4) - E^2(\lambda_{0,1}^2)) E\left(\left(\frac{1}{1+d_i^2}\right)^2\right)$$

$$= \alpha_0^2 p_0^2 (E(\lambda_{0,1}^4) - E^2(\lambda_{0,1}^2)) E\left(\frac{1}{1+d_i^2} - \frac{d_i^2}{(1+d_i^2)^2}\right)$$

$$= \alpha_0^2 p_0^2 \left(E(\lambda_{0,1}^4) - E^2(\lambda_{0,1}^2)\right)(\eta(1) + \eta'(1)), \tag{46}$$

where approximation $(i)$ is due to the following inequality:

$$\left(\sum_{i=1}^{m} \frac{1}{1+d_i^2}\right)^2 \leq m \sum_{i=1}^{m} \left(\frac{1}{1+d_i^2}\right)^2. \tag{47}$$

where the equality holds if all $d_i$'s are equal.

*F. Probability Density Function of* $\text{SINR}_0$

The close-form PDF of SINR is known to be difficult to derive. Fortunately, it can be seen from (21) that the SINR is a summation of many positive terms. Therefore, a Gamma distribution can e used to approximate the SINR according to central limit theorem for causal functions [22] as follows.

$$f_{\text{SINR}_0}(x) = x^{a-1} \frac{e^{-x/b}}{b^a \Gamma(a)}, \forall x > 0, \tag{48}$$

where $a = \text{E}^2(\text{SINR}_0)/\text{Var}(\text{SINR}_0)$, $b = \text{Var}(\text{SINR}_0)/\text{E}(\text{SINR}_0)$, and $\Gamma(a)$ is the gamma function. Then, the CDF (cumulative distribution function) of $\text{SINR}_0$ is

$$F_{\text{SINR}_0}(x) = \int_0^x f_{\text{SINR}_0}(x)dx = \frac{1}{\Gamma(a)} \int_0^{x/b} t^{a-1} e^{-t} dt. \tag{49}$$

In Fig. 2 and Fig. 3, we respectively plot the CDF of SINR with 2 and 20 interfering nodes when there are 4 antennas at each station. Assume that the length of the tagged link, $c_0$, is normalized to 1 and the average received SNR $\alpha_0 p_0 / N_0$ is equal to 20 dB. The interfering links are uniformly distributed within a disc with $R = 3$. The minimum separation between the tagged receiver and interferers is $\varepsilon = 0.1$. From the figures, we can see that although the analytical results come from asymptotic analysis, they match the simulation results very well even with a small number of antennas.

## IV. LINK-LAYER THROUGHPUT CAPACITY

In this section, we investigate the link-layer throughput capacity of wireless ad hoc networks, which is defined as the total data rate that can be successfully delivered through all single-hop links *per unit area*. Assume that an active link transmits at data rate $q$. The transmission is successful only when SINR at the receiver side is higher than a threshold, $\text{SINR}_{th}$, which is a function of $q$. To be more specific, the relationship between $q$ and $\text{SINR}_{th}$ is defined as

$$q = \log_2(1 + \text{SINR}_{th}). \tag{50}$$

Denote by $P_{out}$ the probability of transmission failure of a link, which is calculated from the CDF of

$\text{SINR}_0$ derived in the last section.

$$P_{out} = \Pr(\text{SINR}_0 < \text{SINR}_{th})$$
$$= F_{\text{SINR}_0}(\text{SINR}_{th}) \tag{51}$$

Therefore, the throughput of a communicating link is given by

$$Th = (1 - P_{out})q. \tag{52}$$

If the radius of the network $R$ is very large so that the edge effect is negligible, we can assume that each link experiences homogeneous channel and interference conditions. As a result, the throughput is the same for all links in the network. When there are $K+1$ active links (one tagged link and $K$ interfering links) simultaneously transmitting in the network, we can evaluate the capacity of the network as the summation of the throughput of all the links.

$$\tilde{C}(K+1) = (K+1)(1 - P_{out})q. \tag{53}$$

In wireless ad hoc networks, the number of active links varies from time to time due to the random-access nature of links. Assume that there are in total $L$ links per unit area and each link transmits with a probability $p_t$. Then, the average number of active links is equal to

$$K_0 = \rho_0 \pi R^2 \tag{54}$$

where $\rho_0 = Lp_t$ is the average number of active links per unit area. When $L$ is large, the number of active links follows Poisson distribution. The probability of having $K+1$ active links in the network is given by

$$\Pr(K+1) = \frac{K_0^{K+1} e^{-K_0}}{(K+1)!}. \tag{55}$$

Finally, we have the link-layer throughput capacity of the network as

$$C = \frac{1}{\pi R^2} \sum_{K=0}^{\infty} \tilde{C}(K+1) \Pr(K+1). \tag{56}$$

## V. SIMULATION AND NUMERICAL RESULTS

As shown in the last section, link-layer capacity of wireless networks heavily depends on the number of simultaneously active links within a unit area. This section investigates the impact of the density of active links on the capacity through numerical results. Moreover, the effect of incomplete channel state information is studied.

Similar to Fig. 2 and Fig. 3, we normalize the length of the tagged link to 1 and assume that the average SNR at the tagged receiver is 20dB. The SNR threshold $SINR_{th}$ for the communication pair is 10dB. For simplicity, assume that all transmitters have the same transmission power. Around the tagged receiver, interfering links are uniformly distributed in the space. The minimum separation between the tagged receiver and interferers is $\varepsilon = 0.1$. Given an average density of active links $\rho_0$, the number of active links is randomly generated according to Poisson distribution in (56).

We first validate the analytical results derived in the previous sections by comparing them with simulation results. In Fig. 4 and Fig. 5, the mean and second moment of SINR are plotted against average link density when there are 2, 4, and 6 antennas at each node, respectively. It is not surprising that both mean and second moment decrease as the link density increases. The figures show that our analytical results match the simulations well.

In Fig. 6, link-layer throughput capacity defined in (56) is plotted against the density of active links, $\rho_0$. From the figure, we can see that when the active-link density is low, capacity increases with the number of active links, as the interference can be well handled by multiple antennas. However, when $\rho_0$ exceeds a certain level, co-channel interference becomes so severe that link-layer capacity starts to decrease. As expected, the optimal density of active links that maximizes link-layer capacity increases with the number of antennas, for more co-channel interference can be tolerated when there are a larger number of antennas at each station. Moreover, link-layer capacity increases as the number of antennas increases. For example, the maximal capacity for networks with 2, 4, and 6 antennas is about 0.25, 0.78, and 1.53 bps/Hz/$m^2$, respectively. A close observation of the figure reveals an interesting fact: The maximal capacity increases faster than the number of antennas. In particular, the *normalized* maximal capacity (normalized by the number of antennas) is equal to 0.125, 0.195, and 0.255 bps/Hz/$m^2$, respectively. This provides a strong incentive in deploying multiple antennas in future wireless networks.

In Fig. 7, the optimal active-link density, denoted by $\rho^*$, is plotted as a function of the number of antennas at each station. To validate the analysis, simulation results are also plotted in the figure. The figure shows that our analysis can accurately predict the optimal density of active links in wireless networks with MIMO links.

In wireless networks, active-link density is directly related to the transmission probability of existing links, as shown in the last section. In traditional wireless networks, transmission probability is usually selected according to the network contention level. In this paper, we argue that the optimal transmission probability should be determined by the characteristics of PHY-layer co-channel interference as well as the interference cancelation capability at each receiver. As Fig. 7 shows, the optimal transmission probability can be accurately calculated through our analysis. The observations in Fig. 6 and Fig. 7 serves as a guideline in designing the transmission probability in wireless networks with MIMO links.

So far, we have assumed CSI at each receiving node. That is, the receiver knows the channel matrices $\hat{\mathbf{h}}_k$ (see eqn. (10)) from all interfering nodes. In practice, however, it is difficult for a receiver to monitor the CSI on all links. Hence, it would be interesting to investigate network capacity in a more practical scenario where only the CSI from neighboring interferers is available. In Fig. 8, we assume that a receiving node only estimates the channel from interferers that are located within distance 2 from the receiver. Interference from other interferers is treated as noise. By restricting the channel-monitoring range, the computational complexity due to channel estimation and MMSE detection can be significantly reduced. The figure shows that the maximum throughput is slightly reduced from $0.8 bps/m^2$ to $0.71 bps/m^2$ when the channel-monitoring range is restricted to 2. Intuitively, the larger the channel-estimation range, the higher the capacity. In real implementation, one can trade off between computational complexity and achievable capacity.

In this paper, we have assumed that the optimal linear detector, MMSE, is deployed at each receiver. In real systems, suboptimal detectors such as zero-forcing (ZF) detector are also widely used due to the easy implementation. In the case of ZF, V in eqn. (13) satisfies

$$\mathbf{V}^* = \mathbf{G}^+, \tag{57}$$

where $\mathbf{G}^+$ denotes the psudo inverse of matrix $\mathbf{G}$. For comparison purpose, we investigate the link-layer capacity when ZF detector is deployed in Fig. 9. Note that the number of interferences a ZF detector can handle is no more than $m-1$, where $m$ is the number of antennas. In the figure, we assume that the strongest $m-1$ interferences are canceled by the ZF detector. Similar to the case of MMSE detector, the figure shows that there exists an optimal active-link density when ZF detector is deployed. However, the maximum capacity is reduced by more than 30% compared with the MMSE detector. Due to the lower interference cancelation capability of ZF compared with MMSE, the optimal link density is also reduced.

VI. CONCLUSION

In this paper, we have investigated the link-layer throughput capacity of wireless ad hoc networks when multiple antennas are deployed at each node. In contrast to previous work where network capacity is calculated as if each link exclusively occupies a geometric area, we have argued that it is indeed the characteristics of PHY-layer interference and the interference cancelation capability of receivers that determines the network capacity. This is especially true in networks with MIMO links, where links can transmit simultaneously in the vicinity of each other, with co-channel interference being reduced via space-domain signal processing. One key contribution of this work is the characterization of distribution of post-detection SINR of MMSE receivers when the number and locations of interferers are *random*. The PHY-layer SINR is then translated into MAC-layer throughput capacity in wireless ad hoc networks. We have shown that there exists an optimal transmission probability that maximizes network throughput capacity. In particular, the optimal transmission probability is determined by the number of antennas as well as the multiuser detection scheme deployed at each node. This observation serves as a guideline for the design of MAC protocols in future wireless ad-hoc networks with MIMO links.

References:

[1]   G. J. Foschini and M. J. Gans, ``On limits of wireless communications in a fding environment when using multiple antennas,'' *Wireless Personal Commun.: Kluwer Academic Press*, no. 6, pp. 311-335, 1998.

[2]   E. Telatar, ``Capacity of multi-antenna Gaussian channels,'' *Eur. Trans. Telecom ETT*, vol. 10,no. 6, pp. 585-596, Nov. 1998.

[3]   Q. H. Spencer, C. B. Peel, A. L. Swindlehurst, and M. Haardt, ``An introduction to the multi-user MIMO downlink,'' *IEEE Commun. Mag.*, pp. 60-67, Oct. 2004.

[4]   G. Caire and S. Shamai, ``On the achievable throughput of a multiantenna Gaussian broadcast channel,'' *IEEE Trans. Inf. Theory*, vol.49, pp. 1691-1706, July 2003.

[5]   P. Viswanath and D. Tse, ``Sum capacity of the vector Gaussian broadcast channel and uplink-downlink duality ,'' *IEEE Trans. Inf. theory*, vol. 49, pp. 1912-1921, Aug. 2003.

[6]   W. Rhee and J. M. Cioffi, ``On the capacity of multiuser wireless channels with multiple antennas,'' *IEEE Trans. Inf. Theory*, vol. 49, pp. 2580-2595, Oct. 2003.

[7]   W. Yu, W. Rhee, S. Boyd, and J. M. Cioffi, ``Iterative water filling for Gaussian vector multiple-access channels,'' *IEEE Trans. Inf. Theory*, vol. 50, no. 1, pp. 145-152, Jan. 2004.

[8]   G. Anastasi, M. Conti, and E. Gregori, *IEEE 802.11 Ad Hoc Networks: Protocols, Performance*

*and Open Issues*, New York: IEEE Press?CWiley, 2004.

[9]     P. Gupta and P. R. Kumar, ``The capacity of wireless networks,'' *IEEE Trans. Inf. Theory*, vol. 46, pp. 388-404, Mar. 2000.

[10]    S. Toumpis and A. J. Goldsmith, ``Capacity regions for wireless ad hoc networks,'' *IEEE Trans. Wireless Commun.*, vol. 2, pp. 736-748, Jul. 2003.

[11]    R. S. Blum, ``MIMO capacity with interfrence,'' *IEEE J. Selected Area Commun.*, vol. 21, no. 5, pp. 793-801, June 2003.

[12]    R. S. Blum, ``On the capacity of cellular systems with MIMO,'' *IEEE Comm. Lett.*, vol. 6, no. 6, pp. 242-244, June, 2002.

[13]    W. Choi and J.G. Andrews, ``On spatial multiplexing in cellular MIMO-CDMA systems with linear receivers,'' in *Proc. IEEE Int. Conf. Commun.*, vol. 4, pp. 2277-2281, May, 2005.

[14]    Y. Tokgoz and B. D. Rao, ``Performance analysis of maximum ratio transmission based multi-celluar MIMO systems,'' *IEEE Trans. On Wireless Comm.*, vol. 5, no. 1, pp. 83-89, Jan. 2006.

[15]    B. Chen and M. J. Gans, ``MIMO communications in Ad Hoc networks,'' *IEEE Trans. on Sig. Processing*, vol. 54, no. 7, pp. 2773-2783, June 2006.

[16]    M. Zorzi, J. Zeidler, A. Anderson, B. Rao, J. Proakis, A. L. Swindlehurst and M. Jensen, ``Cross-layer issues in MAC protocol design for MIMO ad hoc networks,'' *IEEE Wireless Comm.,* vol. 13, no. 4, pp.62-76, Aug. 2006.

[17]    P. A. Dighe, R. K. Mallik, and S. S. Jamuar, ``Analysis of transmit receive diversity in Rayleigh fading,'' *IEEE Trans. Commun.*, vol. 51, no. 4, pp. 694-703, Apr. 2003.

[18]    A. M. Tulino and S. Verdu, *Random Matrix Theory and Wireless Communications*, Delft : Now, 2004.

[19]    P. Li, D. Paul, R. Narasimhan, and J. Cioffi, ``On the distribution of SINR for the MMSE MIMO receiver and performance analysis,'' *IEEE Trans. on Inf. Theory*, Vol. 52, No. 1, pp. 271-286, Jan. 2006.

[20]    R. J. Muirhead, *Aspects of Multivariate Statistical Theory*. Wiley, 1982.

[21]    N. R. Goodman, ``Statistical analysis based on a certain multivariate complex gaussian distribution (an introduction),'' *Annals of Mathematical Statistics*, vol. 34, pp. 152-177, 1963.

[22]    A. Papoulis, *The Fourier Integral and its Applications*. New York: McGraw-Hill, 1962.

[23]    J. Zhang and S. C. Liew, ``Capacity improvement of wireless ad hoc networks with directional antennae,'' *ACM MobiCom'05*, Aug. 2005.

[24]    J. Zhang and S. C. Liew, ``Capacity improvement of wireless ad hoc networks with directional

antennae," *IEEE VTC'06*, vol. 2, pp. 911-915, 2006.

[25]   S. Verdu, *Multiuser Detection*, Cambridge University Press, Cambridge, UK, 1998.

[26]   H. V. Poor and S. Verdu, ``Probability of error in MMSE multiuser detection," *IEEE Trans. Inf. Theory*, vol. 43, no. 3, pp. 858-871, May 1997.

[27]   D. N. C. Tse and O. Zeitouni, ``Linear multiuser receivers in random environments," *IEEE Trans. Inf. Theory*, vol. 46, no. 1, pp. 171-188, Jan. 2000.

[28]   D. Guo, S. Verdu, and L. K. Rasmussen, ``Asymptotic normality of linear multiuser receiver outputs," IEEE Trans. Inf. Theory, vol. 48, no. 12, pp. 3080-3095, Dec. 2002.

[29]   G. G. Raleigh and J. M. Cioffi, "Spatio-temporal coding for wireless communications," IEEE Trans. Commun., vol. 46, pp. 357-366, March 1998.

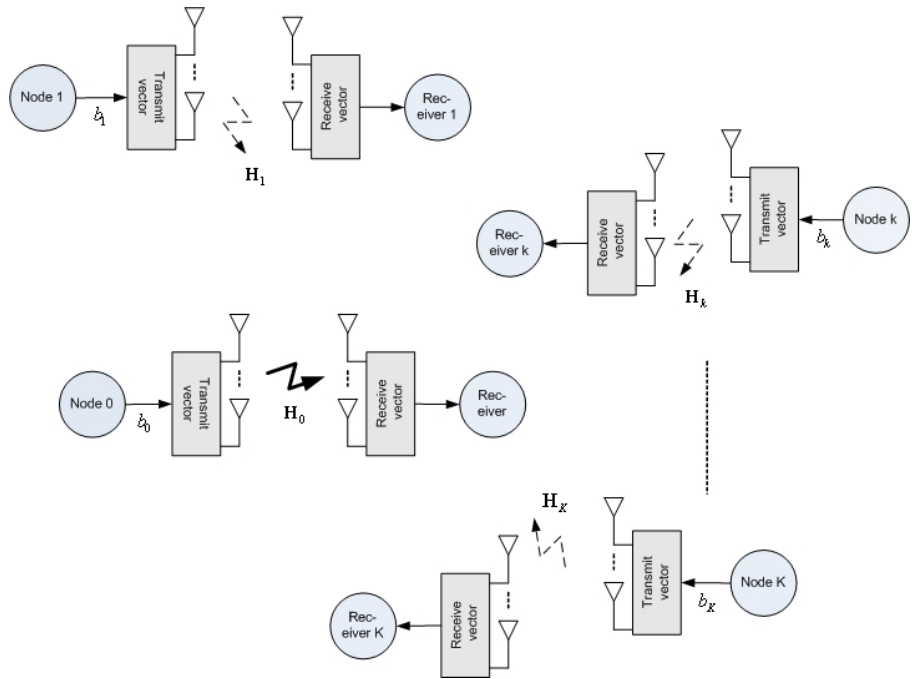

Fig. 1: System model

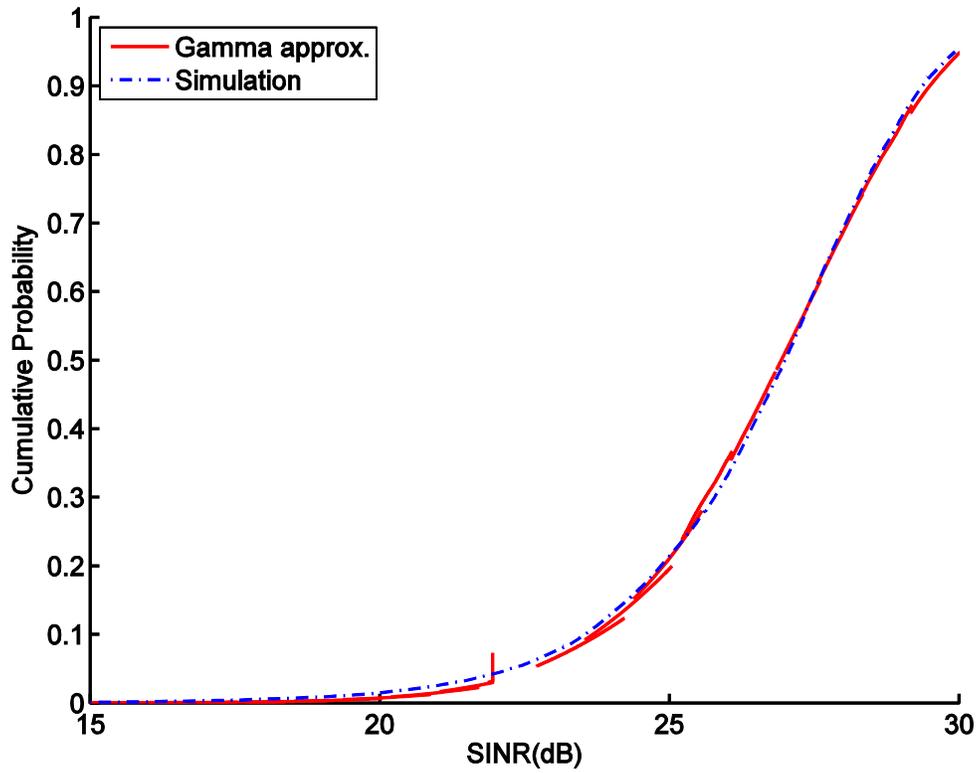

Fig. 2: CDF of SINR when there are 2 interfering nodes

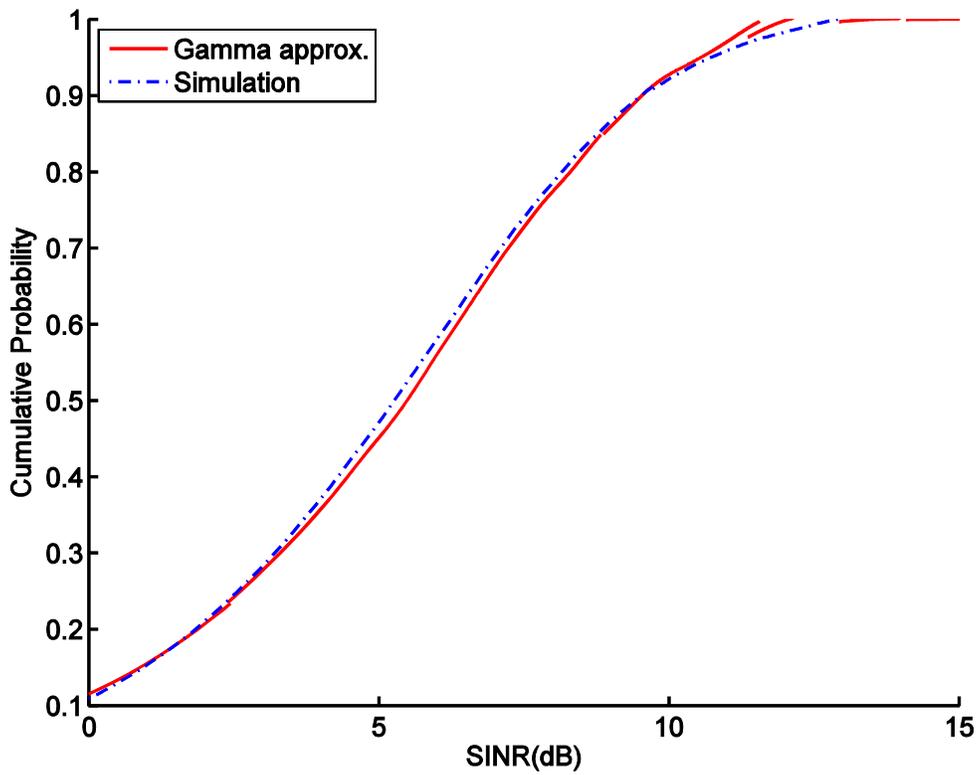

Fig. 3: CDF of SINR when there are 20 interfering nodes

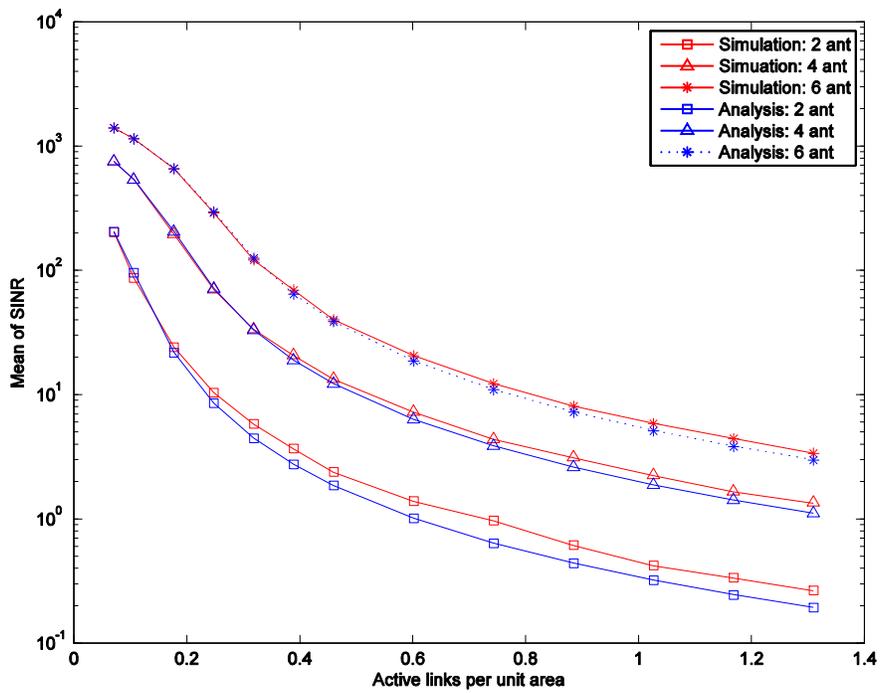

Fig. 4: Mean of SINR

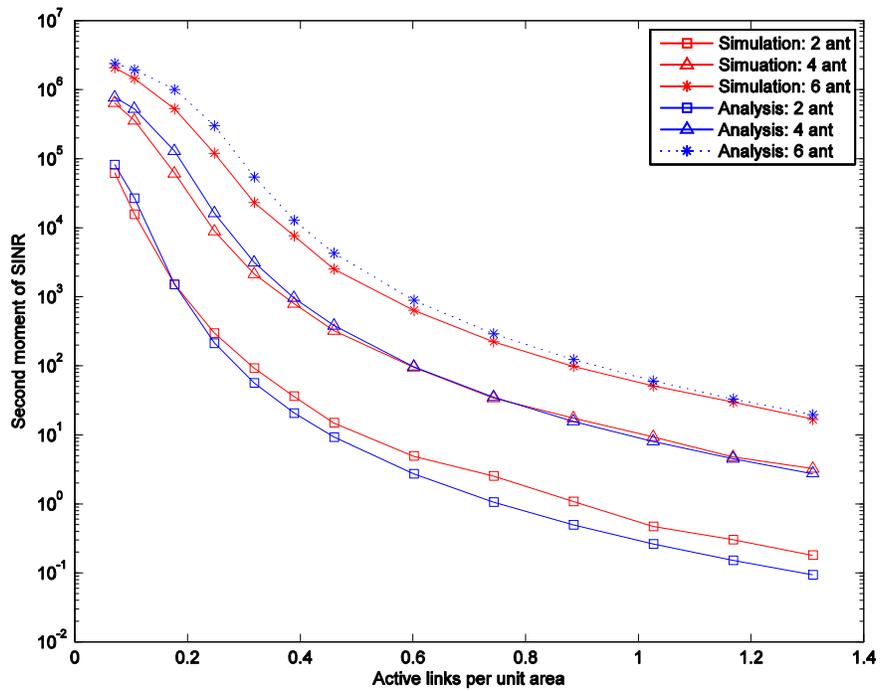

Fig. 5: Second moment of SINR

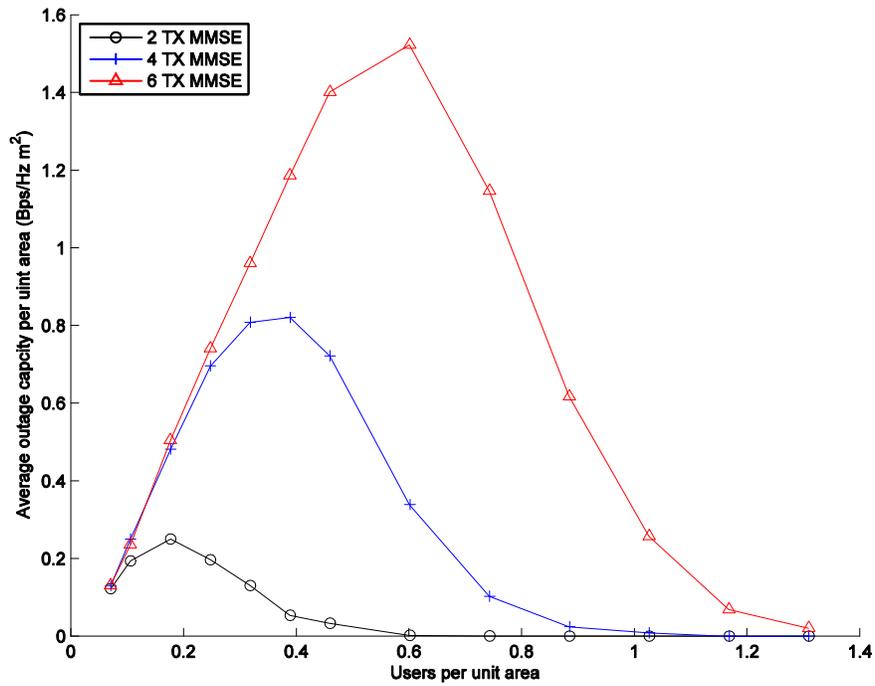

Fig. 6: Throughput capacity vs. the density of interfering nodes

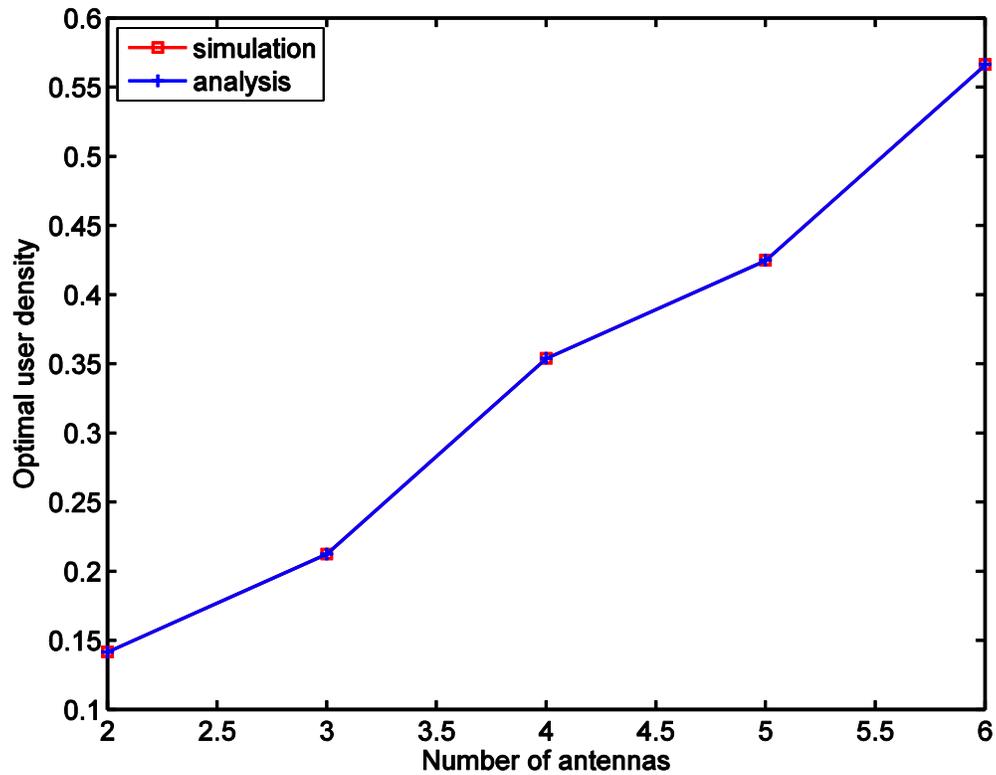

Fig. 7: Optimal active-link density

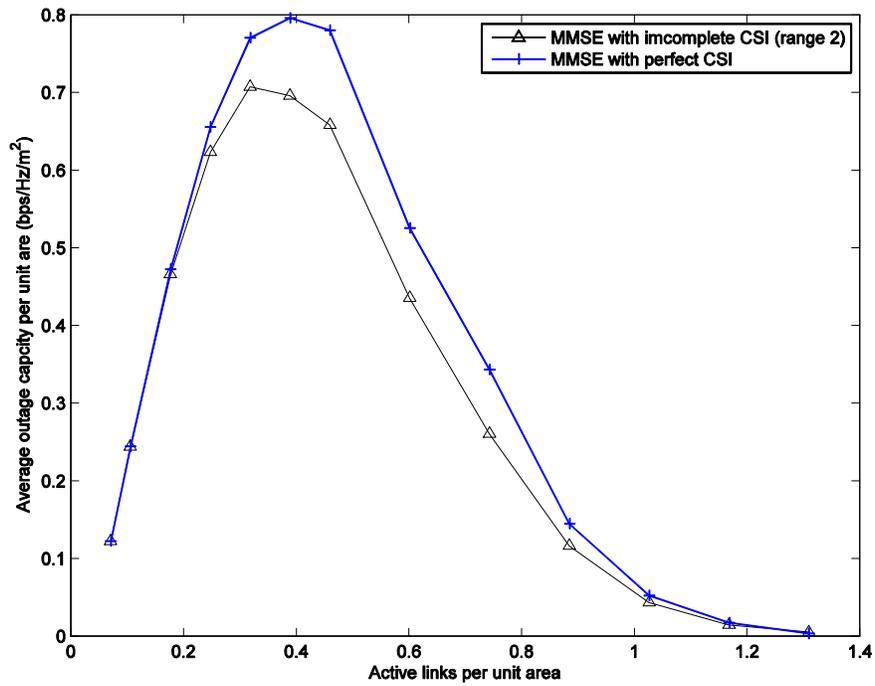

Fig. 8: Throughput capacity with incomplete channel state information

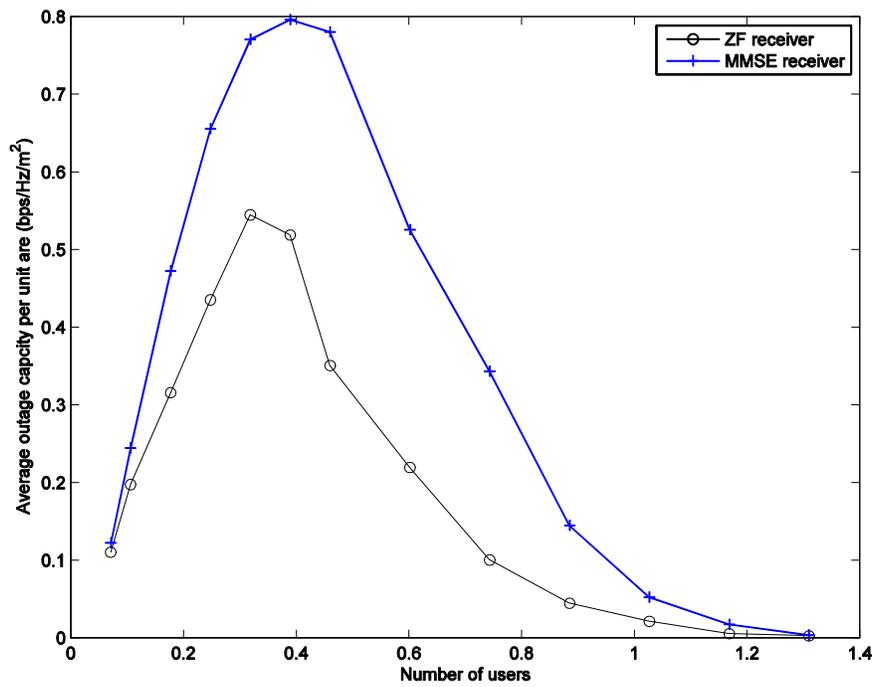

Fig. 9: Throughput capacity of ZF receivers